\documentclass[aps,twocolumn,prd,showpacs,showkeys,preprintnumbers,superscriptaddress,nobibnotes,floatfix,longbibliography,notitlepage,nofootinbib]{revtex4-1}

\pdfoutput=1

\def\apj{ApJ}%
\def\apjs{ApJS}%
\def\apss{Ap\&SS}%
\def\aap{A\&A}%

\usepackage{amsmath}
\usepackage{amsfonts}
\usepackage{amssymb}
\usepackage{mathrsfs}
\usepackage{graphicx}
\usepackage{xcolor}
\usepackage{xspace}
\usepackage{placeins}
\usepackage{booktabs}

\newcommand{\mesa}{\texttt{MESA}}

\newcommand{\abs}[1]{\left| #1 \right|} 
\newcommand\colvec[3][]{\begin{pmatrix}\ifx\relax#1\relax\else#1\\\fi#2\\#3\end{pmatrix}}
\definecolor{darkmagenta}{rgb}{0.55, 0.0, 0.55}
\newcommand{\beq}{\begin{equation}}
\newcommand{\beqn}{\begin{eqnarray}}
\newcommand{\eeq}{\end{equation}}
\newcommand{\eeqn}{\end{eqnarray}}
\newcommand\numberthis{\addtocounter{equation}{1}\tag{\theequation}}
\newcommand{\dbar}{\ensuremath{\mathchar'26\mkern-12mu d}}

\usepackage{array}
\usepackage{bigints}
\usepackage{tabularx}

\usepackage[colorlinks]{hyperref}
\hypersetup{
    colorlinks,
    linkcolor={red!50!black},
    citecolor={blue!50!black},
    urlcolor={blue!80!black}
}
\usepackage{multirow}
\usepackage{upgreek}
\usepackage[capitalise]{cleveref}
\usepackage{soul}

\def\apj{\rm{ApJ}}         
\def\apjs{\rm{ApJS}}        
\def\apss{\rm{Ap\&SS}}       
\def\aap{\rm{A\&A}}        

\begin{document}
\title{New bounds on light millicharged particles from the tip of the red-giant branch}

\author{Audrey Fung}
\affiliation{Department of Physics, Engineering Physics and Astronomy, Queen's University, Kingston ON K7L 3N6, Canada}
\affiliation{Arthur B. McDonald Canadian Astroparticle Physics Research Institute, Kingston ON K7L 3N6, Canada}
\author{Saniya Heeba}
\affiliation{Department of Physics \& Trottier Space Institute, McGill University, Montr\'{e}al, QC H3A 2T8, Canada}
\author{Qinrui Liu}
\affiliation{Department of Physics, Engineering Physics and Astronomy, Queen's University, Kingston ON K7L 3N6, Canada}
\affiliation{Arthur B. McDonald Canadian Astroparticle Physics Research Institute, Kingston ON K7L 3N6, Canada}
\affiliation{Perimeter Institute for Theoretical Physics, Waterloo ON N2L 2Y5, Canada}
\author{Varun Muralidharan}
\affiliation{Department of Physics \& Trottier Space Institute, McGill University, Montr\'{e}al, QC H3A 2T8, Canada} 
\affiliation{Department of Physics, Indian Institute of Technology Kanpur, Kanpur 208016, India}
\author{Katelin Schutz}
\affiliation{Department of Physics \& Trottier Space Institute, McGill University, Montr\'{e}al, QC H3A 2T8, Canada}
\author{Aaron C. Vincent}
\affiliation{Department of Physics, Engineering Physics and Astronomy, Queen's University, Kingston ON K7L 3N6, Canada}
\affiliation{Arthur B. McDonald Canadian Astroparticle Physics Research Institute, Kingston ON K7L 3N6, Canada}
\affiliation{Perimeter Institute for Theoretical Physics, Waterloo ON N2L 2Y5, Canada}

\begin{abstract}
\noindent Stellar energy loss is a sensitive probe of light, weakly coupled dark sectors, including ones containing millicharged particles (MCPs). The emission of MCPs can affect stellar evolution, and therefore can alter the observed properties of stellar populations. In this work, we improve upon the accuracy of existing stellar limits on MCPs by self-consistently modelling (1) the MCP emission rate, accounting for all relevant in-medium effects and production channels, and (2) the evolution of stellar interiors (including backreactions from MCP emission) using the \texttt{MESA} stellar evolution code. We find MCP emission leads to significant brightening of the tip of the red-giant branch. Based on photometric observations of 15 globular clusters whose bolometric magnitudes are inferred using parallaxes from \emph{Gaia} astrometry, we obtain robust bounds on the existence of MCPs with masses below 100 keV. 
\end{abstract}

\maketitle

\section{Introduction}
\label{sec:intro}
Stellar interiors are among the best places to probe weakly coupled extensions of the Standard Model (SM), including hidden sectors containing light degrees of freedom~\cite{raffelt1996stars}. Particles within hidden sectors can be produced from extremely rare processes in the thermal plasma of the stellar interior. Despite the weak couplings, the total integrated emission rates can be relatively large due to the high density, temperature, and volume of stars. Highly interactive particles, for instance photons from the SM, have a short mean free path in stellar interiors and diffuse out slowly over tens of thousands of years. In contrast, weakly coupled hidden-sector particles have a long mean free path in the SM plasma, traversing the star within a matter of seconds. Their unimpeded emission can therefore be an important source of stellar energy transport and loss, and the emission of these particles can affect stellar evolution in analogy to that of SM neutrinos. These considerations have provided some of the strongest constraints on low-mass axions, dark photons, scalars coupling to various fermions, millicharged particles (MCPs), and other particles beyond the SM~\cite{davidson1991limits, Davidson:1993sj, blinnikov1994cooling,Davidson:2000hf, Vogel:2013raa, Hardy:2016kme, An:2013yfc, Gondolo:2008dd, Vinyoles:2015aba, Ayala:2014pea,    raffelt1990core,Haft:1993jt,Viaux:2013hca,Viaux:2013lha, Arceo-Diaz:2015pva,MillerBertolami:2014rka, Isern:2008nt,Corsico:2012ki}. 

\begin{figure}
\hspace{-.5cm}
\includegraphics[width=0.51\textwidth]{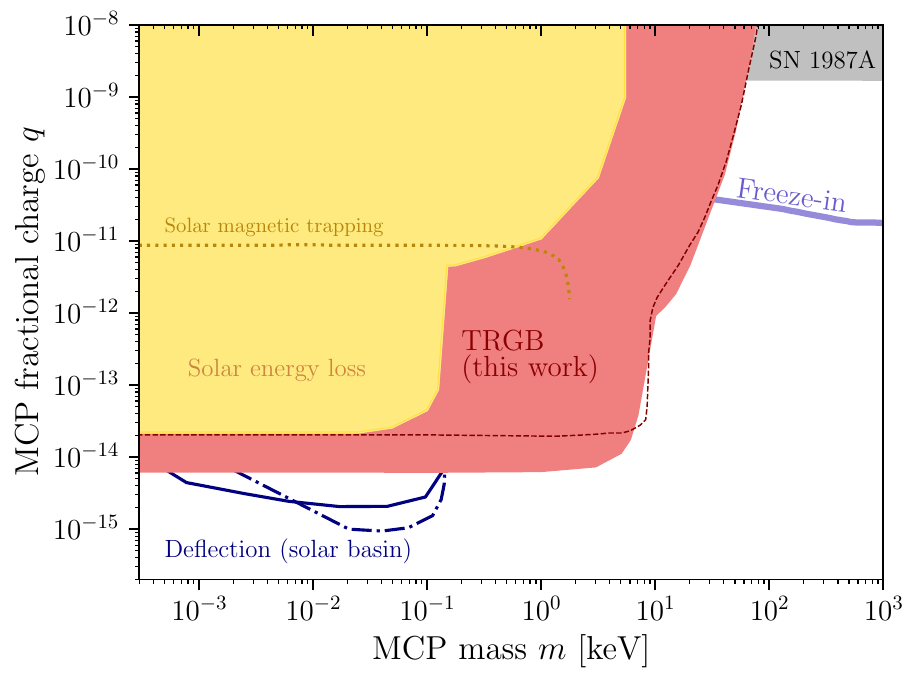}
\vspace{-0.3cm}
    \caption{ 
    Limits on the fractional charge $q$ as a function of mass $m$. The red shaded region shows the limits (at 95\% confidence) obtained in this work while the dashed red line shows previous constraints from the TRGB~\cite{Davidson:2000hf,Vogel:2013raa}. Other shaded regions correspond to existing constraints from Solar energy loss (which may be modified at large values of $q$ due to magnetic trapping)~\cite{Vinyoles:2015khy} and Supernova 1987A~\cite{Chang:2018rso}. Above the purple line, MCPs would be overproduced as dark matter via the freeze-in mechanism~\cite{Dvorkin:2019zdi}. Dark blue lines are the projected sensitivities of a direct deflection setup searching for a solar basin of MCPs with perturbed (solid) and unperturbed (dot-dashed) phase space distributions~\cite{Berlin:2021kcm}.}
    \vspace{-0.5cm}
    \label{fig:constraint}
\end{figure}

Some of the existing stellar limits on hidden sectors rely on a heavily simplified treatment of stellar interiors, in some cases assuming a single plasma frequency $\omega_p$ and temperature $T$ averaged over an entire star or even over an entire population of stars. Some hidden sector emission mechanisms are extremely sensitive to these assumed properties; for instance, even in the SM the neutrino emissivity from plasma processes $Q_\nu$ scales as $Q_\nu\sim \omega_p^{15/2} T^{3/2}$~\cite{Braaten:1993jw}. In this work, we identify MCPs as a particle whose emission from stellar interiors merits a more detailed analysis. Specifically, we consider interactions of the form
\begin{equation}
    \mathcal{L}\supset q e \bar{\chi} \gamma_\mu \chi A^\mu + \bar{\chi}(i \partial -m)\chi
\end{equation}
for Dirac fermion MCPs $\chi$. The fractional MCP charge $q$ could arise in various ways, for instance as the low-energy limit of a theory with an additional light $U(1)'$ gauge field that mixes with SM hypercharge; for the purposes of this work, we treat $q$ and $m$ as phenomenological parameters without specifying their origin. It is timely to more accurately quantify the stellar limit on the existence of MCPs at and above the $\sim$keV mass scale, since MCPs (which would be produced at an irreducible level by the freeze-in mechanism~\cite{Dvorkin:2019zdi}) are a key target of the sub-MeV dark matter direct detection program, with proposed experimental methodologies coming to fruition in the next decades~\cite{Essig:2022dfa}. 

Current stellar limits on MCPs are competitive with cosmological limits on MCPs as dark matter~\cite{Dvorkin:2020xga}. However for $m$ near or above the $\sim$keV scale, the dominant MCP emission channel is a Compton-like process with a rate that is \emph{exponentially} sensitive to the assumed properties of the stellar interior. It is therefore worthwhile to assess the accuracy of the simplifying assumptions that were used in previous analyses. Most notably, the backreaction from the emission of MCPs (which was not considered in previous works) can alter the properties of the stellar interior, potentially amplifying or quenching the effects of MCP emission. A fully self-consistent treatment of the effects of energy loss requires the use of modern stellar evolution codes. 

We focus on the effect of MCPs on the tip of the red-giant (RG) branch (TRGB), which is one of the most sensitive probes of physics in the cores of evolved stars. Our results are summarized in Fig.~\ref{fig:constraint}. Energy loss to neutrinos produced by plasmon decay is known to delay the onset of helium burning in evolved RG stars. Because the TRGB brightness is insensitive to uncertainties in stellar mass, metallicity and modeling details, it is a particularly clean diagnostic of physics beyond the SM. The robustness of the TRGB is what makes it suitable for use as a standard candle in the distance ladder~\cite{Freedman:2019jwv}. In this work, we account for the effects of MCP emission on the TRGB using a modified version of the Modules for Experiments in Stellar Astrophysics (\mesa) code~\cite{Paxton2011, Paxton2013, Paxton2015, Paxton2018, Paxton2019, Jermyn2023}. Using this framework, we simulate stellar evolution varying over MCP parameters and stellar properties like the metallicity. In contrast to previous works that relied on a simple rescaling of the results presented in Ref.~\cite{raffelt1996stars}, we compare the predicted TRGB to a set of 15 recent TRGB measurements from globular clusters (GCs), with distances calibrated astrometrically using \textit{Gaia} parallaxes \cite{Straniero:2020iyi}. 

The rest of this article is organized as follows. In Section~\ref{sec:eloss} we review the computation of energy loss rate from plasmon decay to light particles. We focus on the special case where the MCP is heavier than half the plasma frequency in Section~\ref{sec:offshell}, requiring a proper treatment for the $2\to3$ MCP emission process, $e^- \gamma \to e^- \chi\bar{\chi}$. In Section~\ref{sec:star} we describe the physical effects of MCPs in a RG star and provide details of our implementation of energy loss into the \mesa~stellar evolution code. Our results and constraints are presented in Section~\ref{sec:effect_TRGB}. Concluding remarks and discussion follow in Section~\ref{sec:conclusions}.

\section{Energy loss rates}
\label{sec:eloss}
For a given $m\rightarrow n$ process, the energy density loss rate from a thermal plasma into final state particle $X$ can be expressed as
\begin{align*}
    Q_X &\equiv \frac{d E_X}{dV dt} = \int \prod_{i=1}^m \left(\frac{\dbar^3 p_i}{2 E_i} \right) \prod_{j=1}^n \left(\frac{\dbar^3 p_j}{2 E_j} \right)\\& (2 \pi)^4~ \delta^{(4)}\left( \sum_{i=1}^m p_i - \sum_{j=1}^n p_j\right)  \abs{\mathcal{M}_{m\rightarrow n}}^2 \mathcal{F} E_X \,, \numberthis
    \label{boltzmann}
\end{align*}
where $\mathcal{F}$ is a phase space factor expressed in terms of the phase space density $f$ of each particle,
\beq
\mathcal{F} = \prod_{i=1}^m f_i \times \prod_{j=1}^n (1 \pm f_j ) - \prod_{j=1}^n f_j \times \prod_{i=1}^m (1 \pm f_i ) \,,
\eeq 
and where plus (minus) signs correspond to Bose (Fermi) statistics. Due to the rareness of the processes that produce MCPs, their occupation number in the final state is much smaller order unity, allowing us to approximate $\mathcal{F} \approx \prod_{i=1}^m f_i$ throughout this work (this criterion can be violated in a stellar basin of non-relativistic MCPs~\cite{Berlin:2021kcm}, but here we are concerned primarily with energy loss where the dominant contribution is from the emission of relativistic MCPs). Note that in general, $Q_X$ is a function of the ambient temperature and density; both of these can enter in thermal phase space factors and they additionally lead to non-trivial in-medium effects for non-relativistic plasmas.

\subsection{Emission of low-mass MCPs}
\begin{table*}[t]
\centering
\begin{tabular}{l| p{0.3\textwidth}p{0.35\textwidth}} 
 \hline
 \hline
\textbf{Plasmon Polarization} & 
~\textbf{$\abs{\mathcal{M}_{\gamma^*_{L,T}\rightarrow \chi \bar{\chi}}}^2$} & ~~\textbf{$Q_\chi$}  \\ [0.5ex] 
 \hline 
 & & \\
 \textbf{Transverse} & $8 q^2 e^2 (\omega E_1 - p_1^2 (1-x^2) - k p_1 x) $ & $\frac{2 q^2 e^2}{3 (2 \pi)^3}   \sqrt{1- \frac{4 m^2}{\omega_p^2}} (\omega_p^2 + 2 m^2) \int \frac{ k^2 dk}{1 - e^{\omega/T}} $  \\ 
 & & \\
 \textbf{Longitudinal} & $4 q^2 e^2 \frac{\omega^2}{k^2} (\omega E_1 - 2 E_1^2 + k p_1 x)$ & $\frac{q^2 e^2 \omega_p^2}{3 (2 \pi)^3}  \int k^2 dk  \sqrt{1- \frac{4 m^2}{ \omega_p^2- k^2}}\frac{ \omega_p^2-k^2  + 2 m^2}{\left(\omega_p^2-k^2\right) \left(1 - e^{\omega/T}\right) }$  \\
 & & \\
 \hline \hline
\end{tabular}

\caption{Matrix elements for plasmon decay to MCPs and energy loss rates to MCPs for light MCPs with masses below the plasma frequency (such that plasmon decay is the dominant production channel). The form of the plasmon phase space depends on the dispersion relation. For the transverse mode, $\omega = \sqrt{k^2 + \omega_p^2}$ and no further closed-form analytic progress is possible (note that the integrals for the transverse processes are closely related to the plasmon number density). Meanwhile, the phase space for the longitudinal mode is independent of $k$, since $\omega = \omega_p$, and we integrate only over $k < \sqrt{\omega_p^2 - 4 m^2}$. }
\label{table:1}
\end{table*}

For MCPs that are lighter than the plasma frequency $\omega_p$, the leading production channel is plasmon decay, $\gamma^* \rightarrow \chi \bar{\chi}$, which has no vacuum analog. Plasmons correspond to poles in the in-medium photon propagator, hence their properties depend on the properties of the background. Stellar interiors can be generally characterized as either classical non-relativistic plasmas or degenerate plasmas. We determine plasmon properties using the approximations developed in Ref.~\cite{Braaten:1993jw}. We find that to very good accuracy (better than the few-percent level), for the temperatures and densities relevant to stellar interiors, the residues of the poles in the photon propagator are unity in the relevant classical and degenerate limits. This applies for both transverse modes and longitudinal modes up to a maximum momentum where the plasmon would cross the lightcone, $k_\text{max} \sim \mathcal{O}(\omega_p)$. Furthermore, we approximate the plasma frequency as 
\beq 
\omega_p^2 = \frac{ 4 \pi \alpha n_e}{\sqrt{p_F^2 +m_e^2}} \,,
\label{gen_wp}\eeq 
where $p_F = (3 \pi^2 n_e)^{1/3}$ is the Fermi momentum. In the nondegenerate limit where $p_F \rightarrow 0$, this reduces to the classical result, $\omega_p^2 = 4 \pi \alpha n_e/m_e $. With the plasma frequency of Eq.~\eqref{gen_wp}, we find that to very good approximation the longitudinal and transverse plasmon dispersion relationships are $\omega_L = \omega_p$ and $\omega_T^2 = k^2 + \omega_p^2 $ respectively.

Using the analytic approximations for the plasmon dispersion relations, we can compute the energy loss rate from plasmon decay to two MCPs from Eq.~\eqref{boltzmann}, 
\begin{align*}
\label{eq:Q}
    Q_\chi  = 2 \int & \frac{\dbar^3 p_1}{2 E_1} \frac{\dbar^3 p_2}{2 E_2} \frac{\dbar^3 k}{2 \omega} ~ (2 \pi)^4~ \delta^{(4)}\left( K - p_1 - p_2\right)  \\
    &\times \abs{\mathcal{M}_{\gamma_{L,T}^*\rightarrow \chi \bar{\chi}}}^2 E_1 \,f_*(\omega) \,. 
    \numberthis
\end{align*}
Here, $p_1$ and $p_2$ are the four momenta of final state MCPs, $K = (\omega, \vec{k})$ is the four momentum of the plasmon with $\omega$ and $k$ being linked through the dispersion relations, and the factor of 2 accounts for the energy lost to the second MCP since the integral is identical if we relabel $1\leftrightarrow2$. The matrix elements for the different photon polarizations and MCP spins is in Table~\ref{table:1}, along with the total energy loss rates per unit volume.

\subsection{Emission of high-mass MCPs}
\label{sec:offshell}
\begin{table*}[t]
    \centering
    \begin{tabular}{l| c}
     \hline
     \hline

\textbf{Plasmon Polarization} &  \textbf{$Q_\chi$}  \\ 
 \hline \\
\textbf{Transverse} & $\frac{2 q^2 e^2}{3(2\pi)^3}\int k dk \int \frac{d\omega}{\pi}\rho_T(\omega) f_{\gamma^\ast} \frac{\omega k (\omega^2 - k^2 + 2 m^2)}{\omega^2-k^2} \sqrt{(\omega^2-k^2)(\omega^2-k^2-4m^2)}$ \\ \\
\textbf{Longitudinal} &  $\frac{q^2 e^2}{3(2\pi)^3}\int k dk \int \frac{d\omega}{\pi} \rho_L(\omega) f_{\gamma^\ast} \frac{\omega^3 k (\omega^2 - k^2 + 2 m^2)}{(\omega^2-k^2)^2} \sqrt{(\omega^2-k^2)(\omega^2-k^2-4m^2)}$ \\ \\ 
 \hline \hline      
    \end{tabular}
    \caption{General expression for the energy loss rate to MCPs using the matrix elements provided in Tab.~\ref{table:1} and the plasmon spectral density $\rho_{L,T}(\omega)$ as defined in the text. These expressions reduce to the ones in Tab.~\ref{table:1} in the limit $\omega_p\geq 2m$.}
    \label{tab:offshelldec}
\end{table*}

\begin{figure*}[t]
\includegraphics[width=0.9\textwidth]{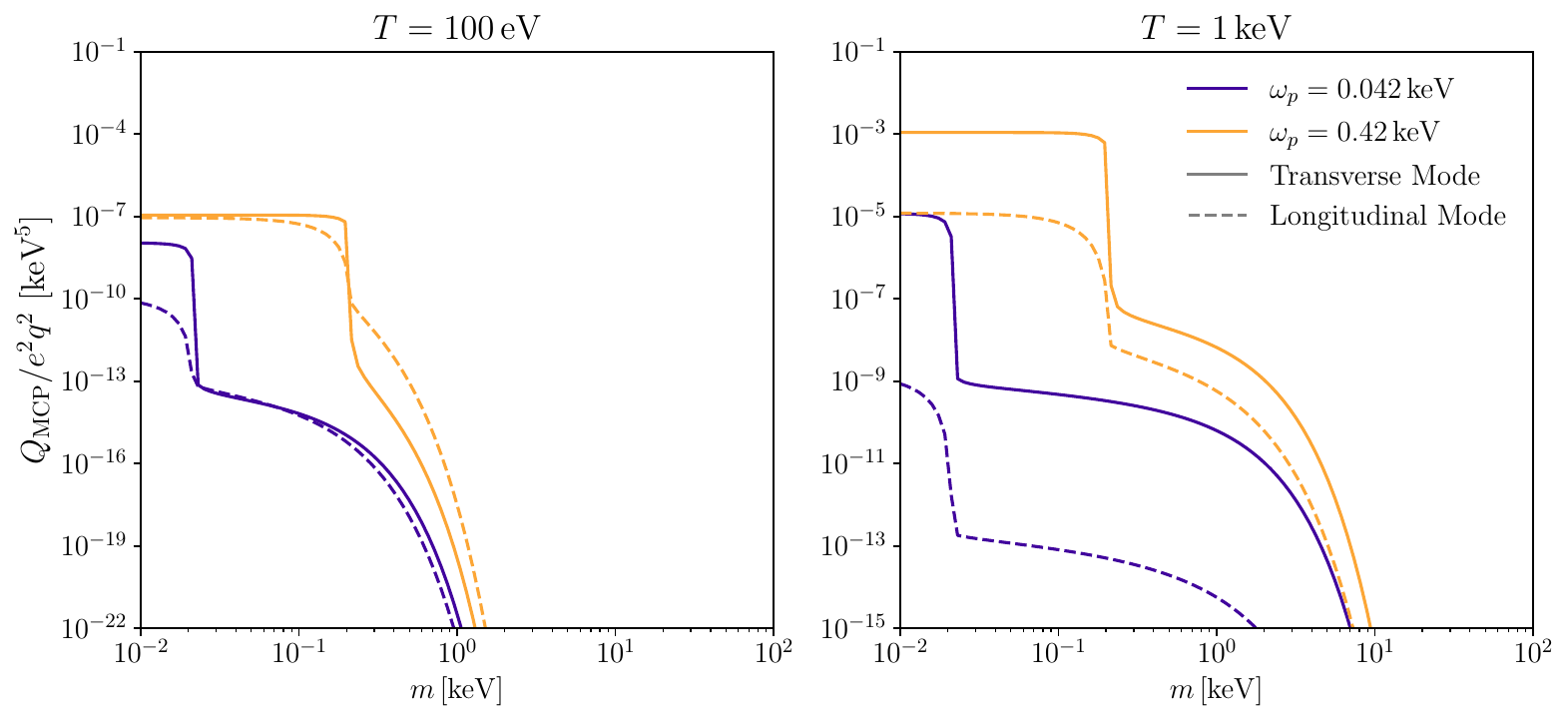}
\vspace{-0.5cm}
\caption{\label{fig:QMCP} Energy loss rate normalised with the  photon-MCP coupling arising from transverse (solid) and longitudinal (dashed) plasmons as a function of the MCP mass for $T=100$~eV (left) and $T= 1$~keV (right), for two representative values of the plasma frequencies, $\omega_p = 0.042$~keV and $\omega_p = 0.42$~keV. These frequencies correspond to $n_e = 10~\mathrm{keV}^3 $ ($1.3\times10^{24}/$cm$^3$) and $n_e = 10^3~\mathrm{keV}^3 $ ($1.3\times10^{26}/$cm$^3$) respectively.}
\end{figure*}

At higher masses, when $\omega_p<2 m$ in a given region of the stellar interior, plasmon decay is no longer kinematically allowed. The next leading order process for producing MCPs is a $2\rightarrow 3$ process resembling Compton scattering but with the final-state photon replaced with an off-shell plasmon that branches into 2 MCPs, $e^-\gamma \to e^- \chi\bar{\chi}$. It has been previously shown that the energy loss through this process can be modelled in terms of the decays of a thermally distributed population of ``off-shell" plasmons \cite{Vogel:2013raa}. 
In fact, Eq.~\eqref{eq:Q} can be modified to account for a population of plasmons with an arbitrary four momentum $K$ as \cite{Vogel:2013raa,Vinyoles:2015khy}

\begin{align*}
\label{eq:Qgen1}
    Q_\chi = 2 &\int  \frac{\dbar^3 p_1}{2 E_1} \frac{\dbar^3 p_2}{2 E_2} \frac{\dbar^3 k}{2 \omega} \frac{d \omega^2}{2\pi} \rho_{L,T}(\omega,\,k)\, E_1 \,f_*(\omega)\ \nonumber\\
    &\times(2 \pi)^4\delta^{(4)}\left( K - p_1 - p_2\right)\abs{\mathcal{M}_{\gamma_{L,T}^*\rightarrow \chi \bar{\chi}}}^2, \numberthis
\end{align*}
where $\rho_{L,T}(\omega, k)$ is the spectral function for the plasmon which can be understood as the probability that a plasmon with the four momentum $K$ exists in the plasma~\cite{Weldon:1983jn},
\begin{align}
\label{eq:spectral}
    \rho_{L,T}(\omega, k) = \frac{2\mathrm{Im}\,\Pi_{L,T}}{(\omega^2 - k^2  - \mathrm{Re}\, \Pi_{L,T})^2 - \mathrm{Im}\,\Pi_{L,T}^2} \,.
\end{align}
Here, $\Pi_{L,T}$ is plasmon self-energy at finite temperature and density. The real part of $\Pi_{L,T}$ sets the dispersion relation of the associated mode, given by the approximations of the previous Subsection. On the other hand, the imaginary part of the self-energy specifies the damping rate. At the energies we consider, the transverse mode is primarily damped through Thomson scattering,  
\begin{align}
    \mathrm{Im}\,\Pi_{T} = \omega n_e\sigma_T = \omega n_e \frac{8 \pi \alpha^2}{3 m_e^2}\,.
\end{align}

In contrast, the longitudinal mode is either Landau damped (when $k> k_\mathrm{max}$), or damped by Thomson scattering (when $k < k_\mathrm{max}$) \cite{Braaten:1993jw,Redondo:2008aa}. Since we only integrate over momenta $k<k_\text{max}$, the damping rate for longitudinal plasmons is defined by the corresponding Thomson scattering cross-section \cite{Redondo:2008aa} 
\begin{align}
    \mathrm{Im}\,\Pi_{L} = \frac{8\pi\alpha^2 n_e}{9 m_e T}\frac{\omega k}{\omega_p}\,.
\end{align}
Using these expressions for the real and imaginary part of the plasmon self-energy and the matrix element for plasmon decay, we can write a general energy loss rate for transverse and longitudinal plasmons into MCPs with an arbitrary mass. 

Integrating Eq.~\eqref{eq:Qgen1} over the MCP phase space, we obtain,
\begin{align*}
        Q_\chi \!=\! \int 
        \!k^2 dk \int_{2 m_\chi}^\infty \frac{\omega d\omega}{\pi} \rho_{L,T}(\omega) \left<\left|M_{\gamma^{\ast}_{L,T}\to\chi \bar{\chi}}\right|^2 \right > f_{\gamma^\ast}(\omega)\,,~ \numberthis
        \label{eq:Qgen2}
\end{align*}
where the momentum-averaged matrix element is
\begin{align}    \left<\left|M_{\gamma^{\ast}_{L,T}\to\chi \bar{\chi}}\right|^2 \right > =& \int\frac{\dbar^3 p_1}{2 E_1} \frac{\dbar^3 p_2}{2 E_2} (2 \pi)^4\delta^{(4)}\left( k - p_1 - p_2\right)  \nonumber\\
&\times E_1\abs{\mathcal{M}_{\gamma_{L,T}^*\rightarrow \chi \bar{\chi}}}^2. \numberthis
\end{align}
The spectral function, $\rho_{L,T}(\omega,\,k)$ is highly peaked when the dispersion relation is satisfied, i.e., when the plasmons are on-shell, $\omega^2 - k^2 = \mathrm{Re}\Pi_{L,T}$, where $\mathrm{Re}\,\Pi_{T} = \omega_p^2$ and $\mathrm{Re}\,\Pi_{L} = \omega_p^2 - k^2$. This implies that for $\omega_p \geq 2 m$, the integral over the spectral function is sensitive to this peak and therefore $\int d\omega \rho_{L,T}(\omega,\,k)$ is well approximated by a Dirac delta function. In this case, Eq.~\eqref{eq:Qgen2} reduces to the expressions given in Table \ref{table:1} for both the longitudinal and transverse case. On the other hand, for $\omega_p < 2 m$, the integral over $\omega$ always misses the peak of the spectral function and the energy loss rate is exponentially suppressed. Additionally, it becomes highly sensitive to the MCP mass and the temperature of the stellar interior. This has the potential to significantly impact the shape of the constraint depending on the assumptions about the stellar interior. We provide the complete expressions for $Q_\mathrm{MCP}$ for the longitudinal and transverse case in Table \ref{tab:offshelldec}. 

In Fig. \ref{fig:QMCP}, we show the energy loss rate through the decay longitudinal and transverse plasmon modes as a function of the MCP mass, for $T = 100~\mathrm{eV}$ (left) and $T= 1~\mathrm{keV}$ (right) and for $\omega_p = 0.042\,\mathrm{keV}$ and $\omega_p = 0.42\,\mathrm{keV}$. The energy loss rate is clearly quite sensitive to the temperature, especially for off-shell plasmon decay to high-mass MCPs. Furthermore, for $\omega_p \geq 2 m$, the energy loss rate is independent of the MCP mass, while for $\omega_p \leq 2 m$, it is exponentially suppressed as discussed above. We note that although the transverse mode gives the dominant contribution to the energy loss in a wide range of parameter space, the contribution of the decaying longitudinal mode cannot be ignored and may in fact even exceed the transverse contribution for large masses and low temperatures (left panel of Fig. \ref{fig:QMCP}).

\section{Energy loss in RG stars}
\label{sec:star}
\subsection{Effects of Beyond-SM Particle Emission}
In standard stellar evolution, stars that have exhausted their core supply of hydrogen depart from the Main Sequence. As the inert helium core contracts under gravitational pressure, its internal temperature rises, driving the star upward in luminosity and causing the envelope to expand, cooling the surface of the star, and creating the RG branch in colour-luminosity space. At the same time, a shell surrounding the helium core continues to fuse hydrogen, enlarging the core over time. In low-mass stars, this trend terminates at the TRGB with the ignition of helium, which is almost solely dependent on the mass of the helium core and is largely independent of other stellar properties. 

The required core mass for the onset of helium ignition is rather sensitive to energy loss. For instance, SM plasmon decays to neutrinos have the effect of enlarging the required core mass for helium ignition. Beyond-SM neutrino properties, such as a small charge or magnetic moment, can be constrained because this would yield an even brighter TRGB than observed~\cite{raffelt1990core,Haft:1993jt,Viaux:2013hca,Arceo-Diaz:2015pva,Viaux:2013lha}. Many previous works use a simple criterion in setting limits on physics beyond the SM, demanding that the total energy loss to new states not exceed twice the standard neutrino luminosity of the core (as this would cause the core mass at helium ignition to change by more than 5\% with respect to the standard theoretical expectation~\cite{raffelt1996stars}). However, later work has included self-consistent stellar simulations with new physics. For instance, Ref.~\cite{Arceo-Diaz:2015pva} simulated a population of stars, constraining the neutrino dipole moment based on the TRGB of the $\omega$-Cen GC. More recently, Refs.~\cite{Dolan:2022kul,Dolan:2023cjs} used \mesa~ to simulate the evolution of GCs in the presence of axion-like particles and dark photons, respectively. They found that the more realistic treatment of energy loss can lead to bounds that are stronger than previously-published results for some masses, while opening up other areas of parameter space. There have also been recent developments on the observational side, for example Ref.~\cite{Capozzi:2020cbu} compared a set of TRGB calibrations from other galaxies to local GCs, updating I-band magnitudes using Gaia distances and combining observational and theoretical uncertainties in order to set limits on axion and neutrino properties.

Prior constraints on light MCPs from stellar energy loss have assumed an energy-loss mechanism similar to plasmon decay to neutrinos \cite{davidson1991limits}, i.e. leading to core cooling and a delayed (and therefore brighter) TRGB. 
Ref.~\cite{Vogel:2013raa} updated the computed emission rates by including off-shell plasmon decay. However, constraints were set by a simple rescaling of the limits in Ref.~\cite{raffelt1996stars}, assuming a single plasma frequency across an entire star. Ref.~\cite{Vinyoles:2015khy} computed the effect of MCP emission in the Sun using the Garching Stellar Evolution Code (GARSTEC)~\cite{2008Ap&SS.316...99W} and set relatively robust bounds using helioseismology. In the calculations described below, we aim to apply many of these state-of-the-art developments in the field in order to modernize the bounds on MCPs from the TRGB.

\subsection{Numerical modeling}
\begin{figure}[t]
    \centering
    \includegraphics[width=0.485\textwidth]{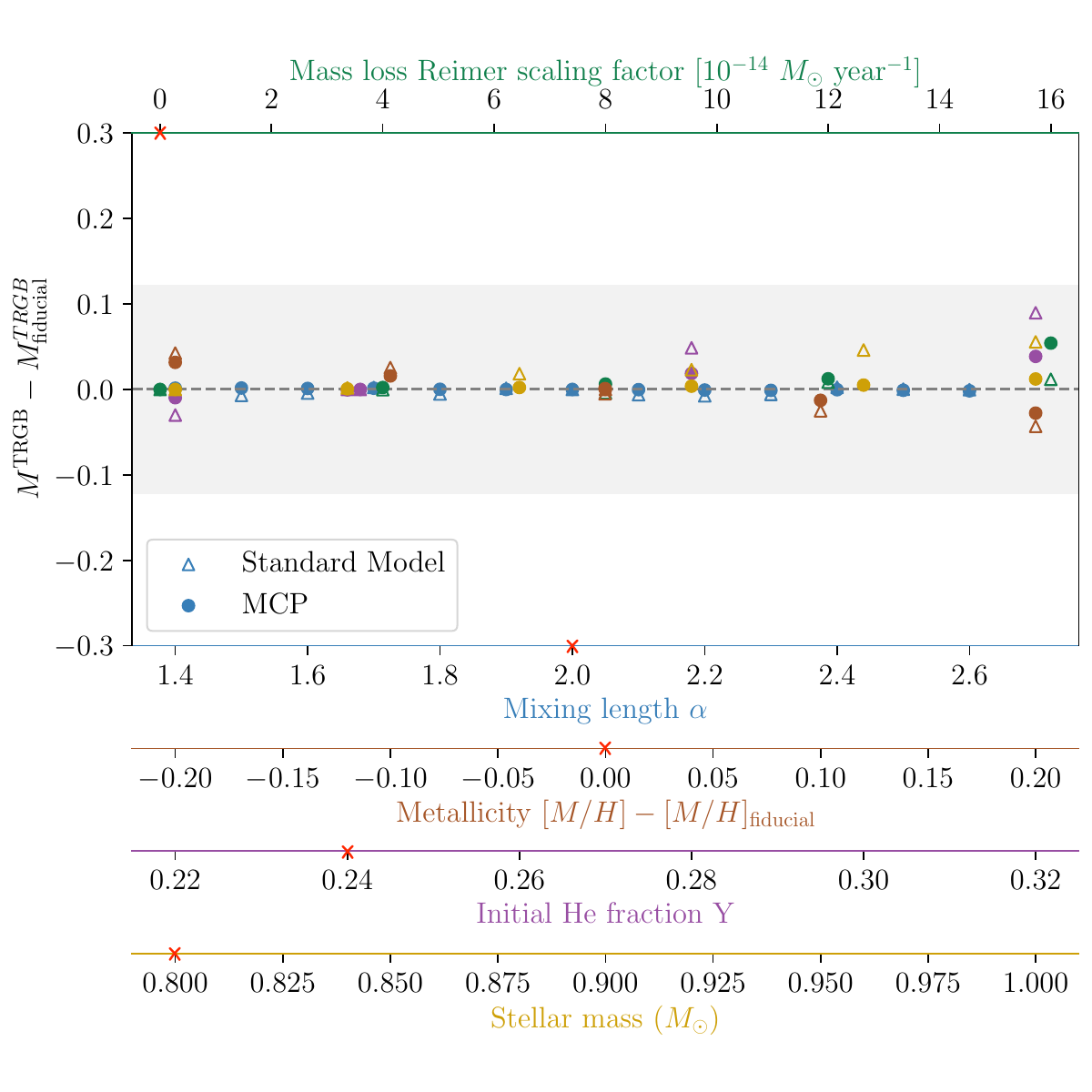}
    \vspace{-0.8cm}
    \caption{Variation of the bolometric magnitude $M^{\mathrm{TRGB}}$ of $\omega$-Cen without (triangles) and with (dots) MCPs ($m = 10$ eV and $q = 2\times 10^{-14}$), compared to our fiducial model, when varying the convection mixing length parameter $\alpha$, metallicity [M/H], initial helium fraction $Y$, and stellar mass. In each set of simulations, only one parameter varies at a time while others are held at their fiducial values (red $\times$ on the corresponding axes). The [M/H] axis covers the range of uncertainty in the metallicity measurement of $\omega$-Cen. The 1$\sigma$ uncertainty on $M^\mathrm{TRGB}$ is indicated by the shaded band.}
\vspace{-0.3cm}
    \label{fig:other_phys}
\end{figure}
 
We use \mesa~to account for the cumulative effect of MCP emission over the lifetime of the star. We use the \texttt{run\_star\_extras} module to include the extra energy loss rate from both the transverse and longitudinal modes as given in Table \ref{table:1} at each time step, based on the stellar temperature and composition profiles. This is then fed into the stellar evolution code via the \texttt{extra\_heat} hook, as a radial array of energies that contribute to the total energy budget of the star. We run our simulations starting with a pre-main-sequence model, and evolve them through to the horizontal branch phase. We take the maximum brightness as the $L^\mathrm{TRGB}$. The absolute magnitude $M_\mathrm{bol}$ is then obtained via the standard definition 
\begin{equation}
    M_\mathrm{bol} - M_\odot = -2.5 \log({L^\mathrm{TRGB}/L_\odot})
\end{equation}
where $M$ and $L$ are absolute magnitudes and luminosities, respectively. 

The GC sample that we use contains clusters with ages $\sim 10$--$13$ Gyr. In this range, the TRGB is represented by stars with masses in the range $0.8$--$0.9$~$M_\odot$. Rather than simulate a full population for each cluster, we perform simulations using a set of fiducial values for the astrophysical and modelling parameters. We additionally vary these parameters to quantify the extent to which they affect our results. As a benchmark, we focus on a 0.8 $M_\odot$ star  and set the initial helium fraction to $Y = 0.241$, the mixing length parameter $\alpha = 2.0$ and we do not include mass loss. Throughout this work, all simulations shown assumed this same set of fiducial values unless otherwise specified. Because GCs cover a wide range of metallicities, we do vary [M/H] to cover the full range spanned by the data, and interpolate results to match the reported value of [M/H] for each cluster. 

To verify the robustness of these assumptions, we have performed a set of simulations with and without the effects of plasmon decay to MCPs that span a range in mass-loss rate, mixing length parameter $\alpha$, metallicity and initial helium fraction. The results are presented in Fig. \ref{fig:other_phys} in terms of the variation of $M^{\mathrm{TRGB}}$ with respect to the fiducial magnitude (when all the parameters are set to their default values). In Fig. \ref{fig:other_phys}, the fiducial value for [M/H] is set to $-1.42$, the value reported for $\omega$-Cen \cite{Straniero:2020iyi}. We also show the 68\% containment on $\omega$-Cen's TRGB luminosity as a grey band. 

We vary the mass-loss rate by scaling the Reimers mass-loss factor $\eta$ from zero to $1.6 \times 10^{-13}\, M_\odot$ yr$^{-1}$, which is twice the typically-used value \cite{schroder2005new} of $8 \times 10^{-14}M_\odot$ yr$^{-1}$, found to reproduce the observed envelope mass of RG stars. As shown in Fig. \ref{fig:other_phys}, even unphysically large mass-loss rates do not appreciably change the the TRGB magnitude. 

Similarly, we vary the convective mixing length $\alpha$ and initial helium fraction over a wide range, finding no significant change in $M^{\mathrm{TRGB}}$, except for especially large values of $Y \gtrsim 0.28$ where dimming of the TRGB is observed. Because of the age of GC stars $\sim 13$ Gyr, the initial helium content should be close to primordial $Y_P \simeq 0.24$, so such a large $Y$ would have to have resulted from an unrealistically large core dredge-up after H exhaustion---this is unlikely, as the effects of He dredge-up should be more-than-compensated by diffusion (see e.g. Ref.~\cite{Chaboyer1992dredge}).

Fig. \ref{fig:other_phys} also shows the effect of varying the metallicity within the reported errors for $\omega$-Cen. The small correlation that can be seen is consistent with the expected effect of metallicity on the TRGB, but is small enough within this range to justify using a fixed value of [M/H] for each individual cluster. 

Finally, Fig. \ref{fig:other_phys} shows the change in luminosity of the TRGB with stellar mass, becoming slightly dimmer for SM stars with increased mass; once energy loss from plasmon decay to MCPs is added, the TRGB actually becomes insensitive to the stellar mass.
\begin{figure}[t]
    \centering
    \includegraphics[width=0.5\textwidth]{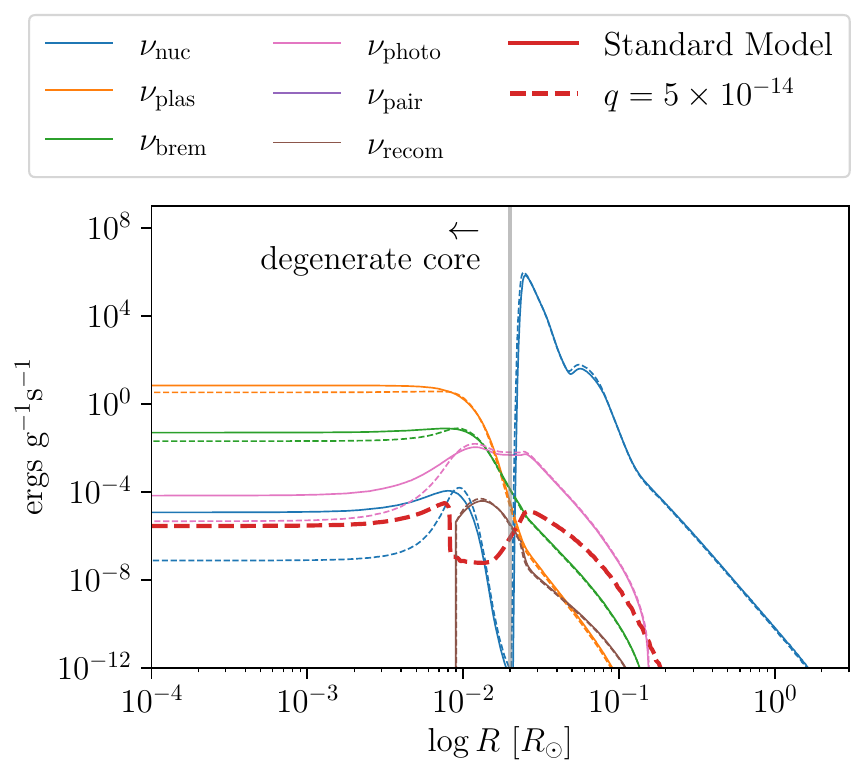}
    \vspace{-0.7cm}
    \caption{Energy loss processes per unit stellar mass in a 0.8~$M_\odot$ star with [M/H] = $-2.05$ just before the TRGB without (solid) and with (dashed) the effects of a light charged particle with $m=7\ \mathrm{keV}$. The self-consistent modeling of MCP emission results in a lower rate of energy loss from SM neutrino processes. MCP energy loss is large outside of the degenerate helium core, in contrast to SM plasmon decay to neutrinos, which occurs in the core.}
    \label{fig:loss}
    \vspace{-0.4cm}
\end{figure}

\begin{figure}[h!]
    \centering
    {\includegraphics[width=0.5\textwidth]{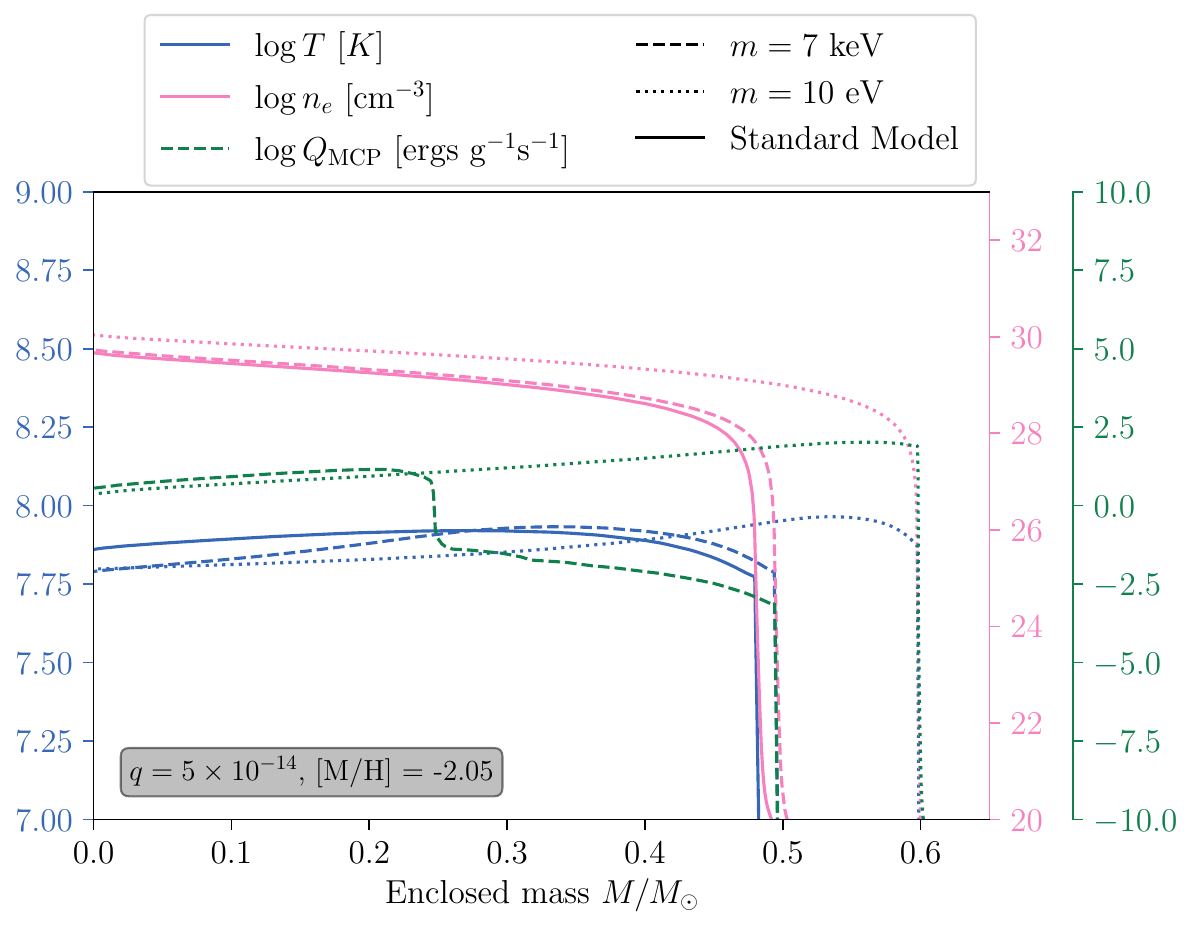}}
    {\includegraphics[width=0.5\textwidth]{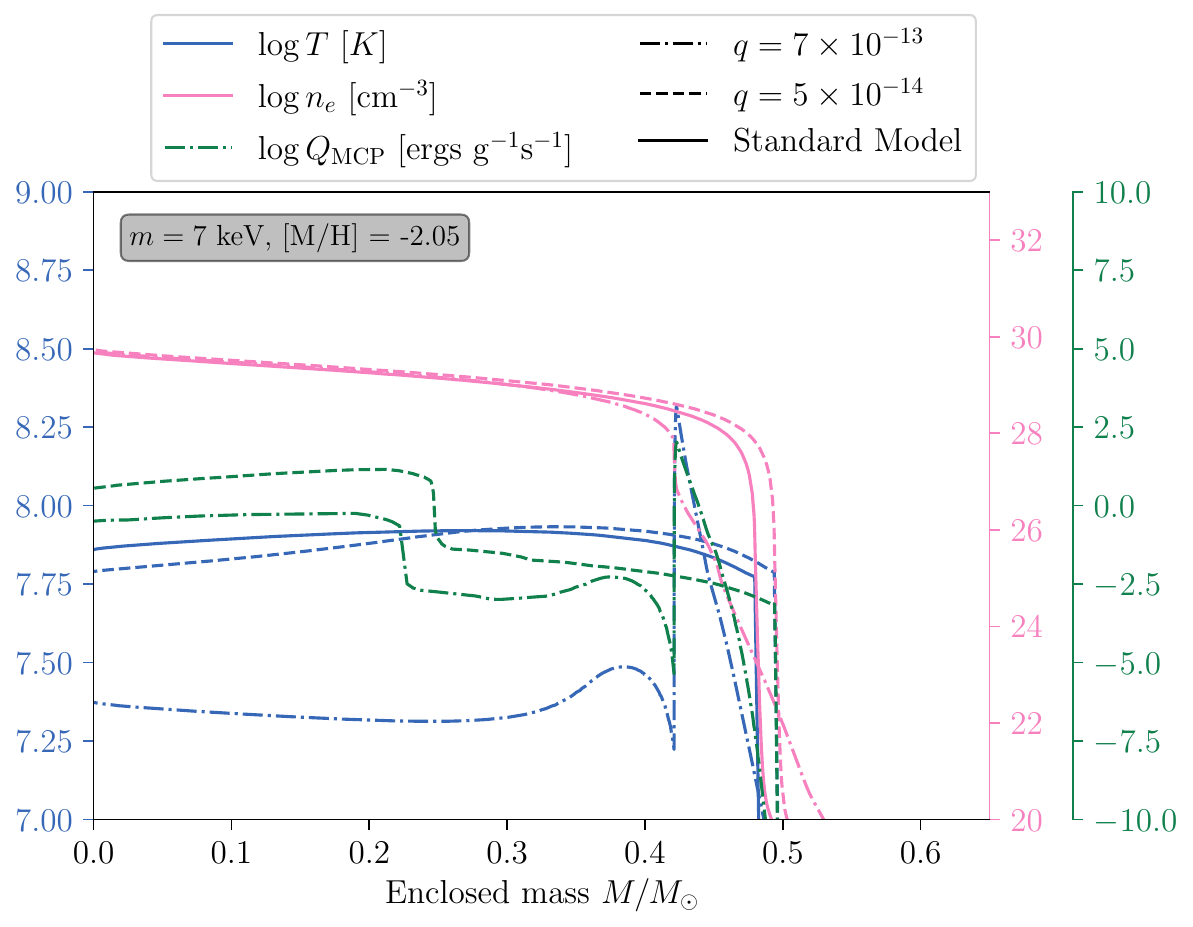}}\\
    {\hspace{-0.33cm}\includegraphics[width=0.5\textwidth]{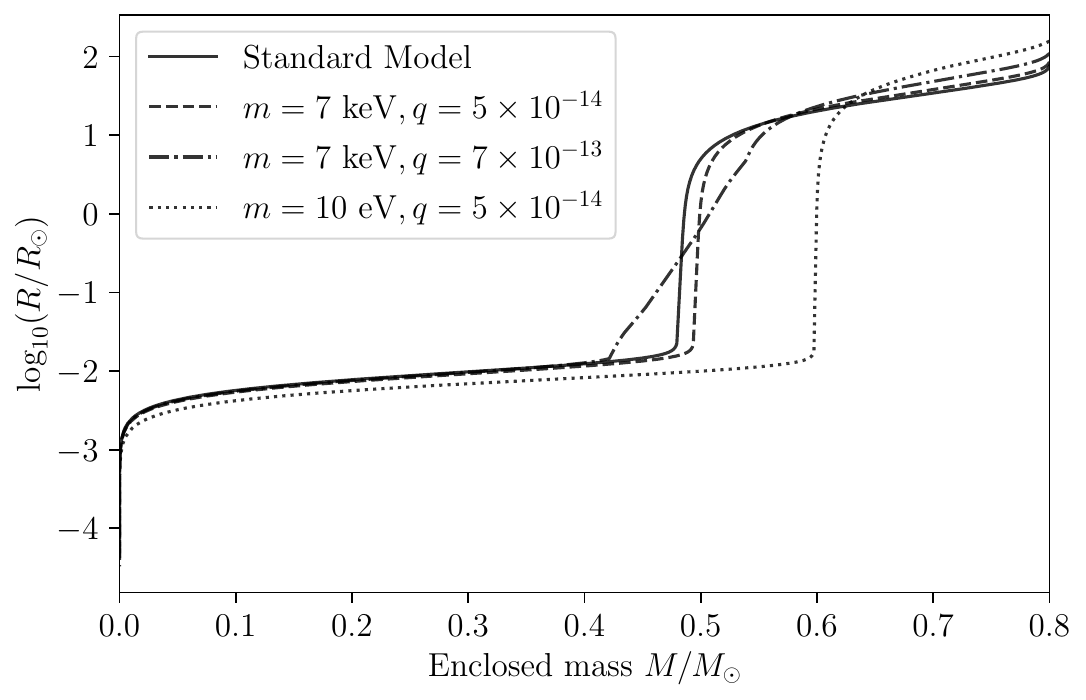}}
    \vspace{-0.6cm}
    \caption{Temperature $T$, free-electron number density $n_e$ and MCP luminosity $Q$ just before the TRGB as a function of enclosed stellar mass with [M/H] = $-2.05$. Top: the MCP charge is fixed at $q = 7\times 10^{-13}$ assuming $m = 7~\mathrm{keV}$~(dashed), $m = 10\ \mathrm{eV}$~(dotted). Middle: the MCP mass is fixed at $m = 7~ \mathrm{keV}$, with $q = 7\times 10^{-13}$~(dot-dashed), $q = 5\times 10^{-14}$~(dashed). In both panels, the SM prediction is shown as a solid line. Bottom: the stellar radius as a function of the enclosed stellar mass.}
    \vspace{-0.5cm}
    \label{fig:core}
\end{figure}

\section{Effect of mcps on the TRGB}\label{sec:effect_TRGB}
\subsection{Simulated stellar interiors}
Fig.~\ref{fig:loss} shows the energy loss profile in a simulated 0.8~$M_\odot$ star due to SM processes and due to the effect of plasmon decay to MCPs, just before the TRGB. The key SM processes include neutrino production from nuclear fusion, plasmon decay, bremsstrahlung, photo-neutrino, pair annihilation, and recombination. The inclusion of MCP emission, which alters the free-electron density and temperature profile of the star shown in Fig.~\ref{fig:core}, also affects the SM energy-loss rates. For all the energy-loss channels, solid lines represent the SM while dashed lines represent the case where the emission of light MCPs with $q = 5 \times 10^{-14}$ is included self-consistently, lowering the energy loss from SM channels. The structure of the emission region for plasmon decay to MCPs, which has prominent support in the envelope outside of the degenerate helium core, also differs substantially from SM plasmon decay to neutrinos which primarily occurs in the core. Thus, the previously used method of modelling MCP production as a scaling correction to the neutrino production rate leads to qualitatively incorrect modelling of the stellar structure. 

Fig.~\ref{fig:core} shows the stellar temperature $T$, free-electron density $n_e$, and MCP emission $Q_\mathrm{MCP}$ profiles as a function of the enclosed mass for evolved stars on the RG branch, just before the TRGB. The lower panel shows the stellar radius as a function of enclosed mass. The stellar profiles are shown for different MCP masses at a fixed charge (top), and for different MCP charges at fixed mass (middle). The solid line represents the stellar structure in the absence of new physics. In all cases, the energy loss from MCPs in the core leads to a reduction in core temperature, a longer period of hydrogen shell burning, and the production of a more massive core before the onset triple-alpha ignition.

 As expected, the effect on the core mass is larger for smaller $m$, where MCP emission in the entire core can occur. The high plasma frequency required for on-shell decay to more massive MCPs, discussed in Section~\ref{sec:offshell}, leads to a suppression of MCP emission, resulting in an additional step down of $Q_\text{MCP}$ in the inner core. The middle panel of Fig.~\ref{fig:core} shows the effects of a larger value of $q$ on a star for $m = 7$~keV. Here, $q$ is increased to $7 \times 10^{-13}$, around an order of magnitude larger than constraints that we obtain at this MCP mass. In this case, the core temperature drops by a factor of three, actually lowering MCP emission from the inner core in comparison to models with a lower charge. The subsequent core contraction ultimately changes the qualitative features of the stellar structure. A shallower density gradient develops, erasing the sharp boundary between the helium core and the envelope, producing a broad, high-temperature shell leading to enhanced MCP energy loss. The main features remain, notably a larger helium region, and a TRGB that is a full 0.8 Mag brighter than in the SM---more than three standard deviations away from the measured value for M92.

\begin{figure}
    \centering
\includegraphics[width=0.48\textwidth]{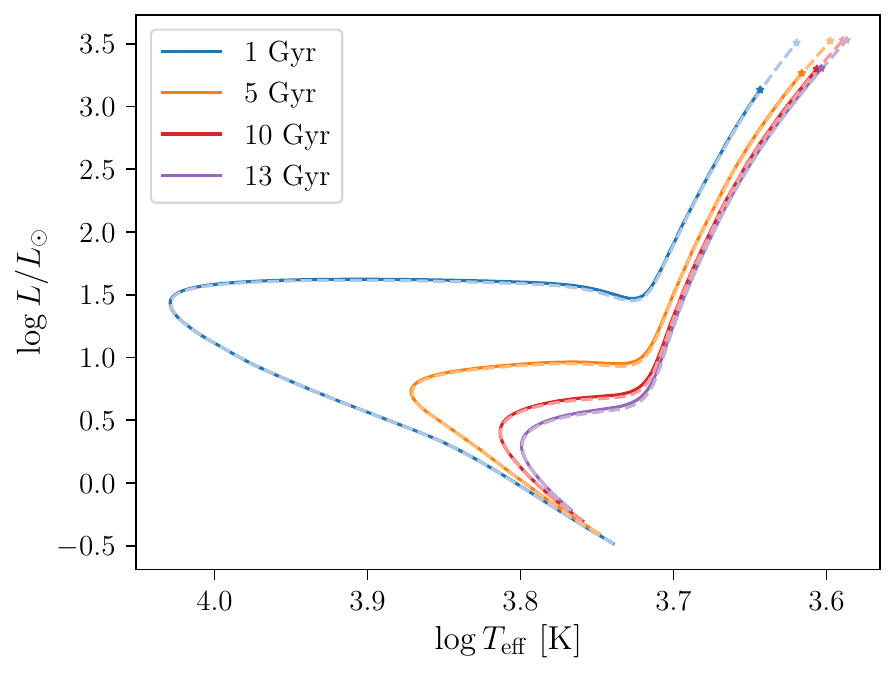}
\vspace{-0.6cm}
    \caption{ Stellar isochrones for a cluster of stars with stellar masses between 0.7~$M_\odot$ and 3~$M_\odot$ and a metallicity representative of $\omega$-Cen, [M/H] = $-1.42$, evolved up to the TRGB (indicated with a star) without MCPs (solid) and with MCPs  with $\textit{m} = 10~\text{eV}$ and $q = 2 \times 10^{-14}$ (dashed). The isochrones were computed using the \mesa~Isochrones and Stellar Tracks package~\cite{2016ApJS..222....8D}. }
    \label{fig:isochrones}
    \vspace{-0.4cm}
\end{figure}

The net effects of MCP emission lead to a delay in helium ignition and thus a higher luminosity (lower bolometric magnitude) as a star reaches the TRGB. We illustrate the effect on a population of stars with isochrones representative of $\omega$-Cen shown in Fig \ref{fig:isochrones}. These colour-luminosity diagrams show the distributions for a population of stars with [M/H] = $-1.42$ and masses between 0.7$-$3~$M_\odot$, taken at age snapshots of 1, 5, 10 and 13 Gyr in the SM (solid lines) and in the presence of a $q= 2 \times 10^{-14}$ MCP in the low-mass $m\ll1$~keV limit (dashed lines). For representative ages of Milky Way GCs ($\sim 10-13$ Gyr), the TRGB luminosity increase due to MCPs remains insensitive to the exact age. 

 \subsection{Constraints}
 \label{sec:constraints}
In order to set limits on the MCP charge $q$ as a function of the MCP mass $m$, we compare the results given by our simulation to recently-reported TRGB magnitude measurements. Ref.~\cite{Straniero:2020iyi} obtained bolometric magnitudes for 22 GCs using photometric data from the Hubble Space Telescope and ground-based optical measurements. They report two distinct values of $M_\mathrm{bol}$ based on different distance-measurement techniques: first via a zero-age horizontal branch (ZAHB) calibration, and second with distances measured using \emph{Gaia} astrometric data~\cite{Gaia:2018ydn} and line-of-sight velocities from ground-based instruments including Keck the Very Large Telescope~\cite{baumgardt2019mean}. Even though parallax distances are only provided for 16 of the 22 clusters, they are reported with higher precision and with errors that are more reliably estimated for each cluster. Though we have not explicitly checked in the present work, it is also possible that the properties of the ZAHB luminosity will be modified by the additional MCP emission considered here, whereas astrometry would be unaffected. Therefore, while we compare constraints on MCPs obtained using both techniques (parallax and ZAHB) in Appendix~\ref{app}, for our main results we use the parallax distances. 

The astrometric data set does contain one outlier, NGC 6553, which has a measured magnitude of $M_{\mathrm{bol} }= -4.76 \pm 0.17$. This is in significant tension with the reported ZAHB measurement of $-3.93 \pm 0.25$ and is 0.7 dex away from any of the other reported magnitudes for other GCs. Here, the emission of MCPs would actually lead to a significantly better fit between theory and observation; however, producing a magnitude this low requires a charge that is well beyond values allowed by the combination of other clusters. In some cases, the changes to the stellar interior induced by such a high MCP charge (and therefore a high MCP emission rate) cause our models to fail to converge. For these reasons, we omit NGC~6553 from our analysis.

\begin{figure}[t!]
    \centering
\includegraphics[width=0.485\textwidth]{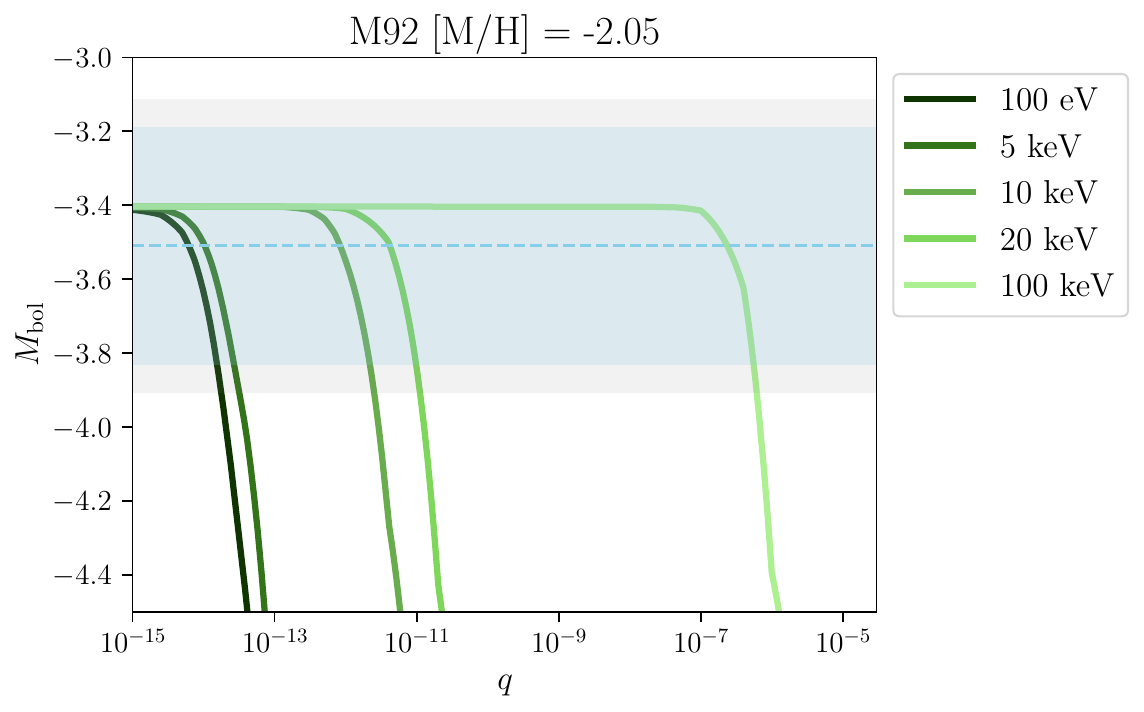}
\vspace{-0.6cm}
    \caption{Bolometric magnitude at the TRGB as a function of MCP charge $q$, for MCP masses at ranging from $100\,\mathrm{eV}$ to 100 keV, from left to right. Fiducial values, marked as red crosses in Fig. \ref{fig:other_phys}, are assumed in all simulations, with [M/H] = $-2.05$, the metallicity of M92. Also shown are the observed magnitude (dashed) and $ 2 \,\sigma$ observational uncertainties of the single cluster M92 on the TRGB luminosity from Ref.~\cite{Straniero:2020iyi} (blue), and total error combined with the theory uncertainty (grey) recommended in Ref.~\cite{2017A&A...606A..33S}.} 
    \vspace{-0.5cm}
    \label{fig:MbolVsQ}
\end{figure}

We simulate stars on a grid of MCP masses, MCP charges, and stellar metallicities covering the range of clusters studied in Ref. \cite{Straniero:2020iyi}. Fig \ref{fig:MbolVsQ} shows the TRGB magnitude predicted in a 0.8 $M_\odot$ star with the metallicity of M92 for a few MCP masses as a function of the fractional charge. This is overlaid with the 68\% containment on the TRGB magnitude of M92 \cite{Straniero:2020iyi}. Once the energy-loss rate becomes large enough, we observe a sharp increase in TRGB luminosity. The impact of increasing $q$ is mitigated at higher MCP masses, as off-shell plasmon decay becomes thermally suppressed, requiring much larger values of $q$ for any effect to be observed.

To evaluate the upper bound on MCP charge at a given MCP mass, we define a Gaussian likelihood $\mathcal{L}(\{M_{i, \text{obs}}\}|q, m) = \prod_i \mathcal{L}_i(M_{i, \text{obs}}|q, m)$ such that 
\begin{equation}
     -2 \sum_i \log \mathcal{L}_i(M_{i, \text{obs}}|q, m)  
    = \sum_i\frac{\left(M_i(q, m)-M_{i, \text{obs}}\right)^2}{\sigma^2_i},
\end{equation}
where the index $i$ runs over the list of GCs,  $M_i(q, m)$ is the predicted bolometric magnitude of the TRGB as a function of MCP charge and mass at the metallicity of a given cluster,  $M_{i, \text{obs}}$ is the bolometric TRGB magnitude inferred from measurement, and $\sigma^2_i$ represents the uncertainties on the inferred bolometric magnitudes. Our uncertainties include the reported observational uncertainties given in Table 6 of Ref.~\cite{Straniero:2020iyi}, added in quadrature to the theoretical uncertainties on the bolometric magnitude predicted by stellar evolution codes. Following the recommendation of Ref.~\cite{2017A&A...606A..33S} we set this theoretical error to 0.12. Applying Wilks' theorem, we use a likelihood ratio test to find the limit on the MCP charge $q_{\text{lim}}$ for for a fixed MCP mass $m$. We set the limit by finding
 \begin{align*}
 2 \Big[\log\mathcal{L}&(\{M_{i, \text{obs}}\}|q_{\text{max}}, m) \numberthis \\
 &- \log\mathcal{L}(\{M_{i, \text{obs}}\}|q_{\text{lim}}, m)\Big] = 2.71 ,
 \end{align*}
where $q_{\text{max}}$ is the value of the MCP charge that maximizes the likelihood at MCP mass $m$ and $2.71$ corresponds to the one-sided 95\% confidence level (CL)~\cite{Cowan:2010js}.

We simulated MCP masses from 10 eV to 100~keV, and show the resulting limits on the charge $q$ in Fig \ref{fig:constraint}. We additionally show prior limits on MCPs from Solar modeling~\cite{Vinyoles:2015khy}, the absence of anomalous cooling of Supernova 1987A~\cite{Chang:2018rso}, and projected constraints from proposed direct deflection searches~\cite{Berlin:2021kcm}. As expected, our constraints plateau for low masses, as $m \ll \omega_p$. For the lowest mass considered here, the resulting 95\% CL limit is
\begin{equation}
    q < 6.3 \times 10^{-15} , 
\end{equation}
a little more than a factor of three stronger than limits based on those of Ref.~\cite{davidson1991limits}. As $m$ increases, the stellar region (notably, the He core) in which on-shell plasmon decay can occur shrinks, leading to a rapid loss of sensitivity above masses corresponding to half the lowest plasma frequency in the core. The ``kink'' seen around 10~keV corresponds to the transition between having kinematically accessible on-shell plasmon decay in some part of the star versus relying purely on off-shell decays throughout the entire star. At high mass, sensitivity is lost at an exponential rate because of the temperature scaling of the off-shell plasmon decay rate. Coincidentally, our limit at high masses is very similar to the estimate performed in Ref. \cite{Vogel:2013raa}, which was obtained by rescaling the low-mass results from Ref.~\cite{davidson1991limits} using a single plasma frequency for a single idealized TRGB star obtained from assuming SM-only properties of the stellar interior. The parts of our improved analysis that would enhance the strength of the constraint (including energy loss from the longitudinal mode, using modern simulations and observations, etc.) are compensated by the effect of MCP emission on the stellar interior, which reduces the energy loss from SM channels. 

\FloatBarrier
\section{Conclusions}
\label{sec:conclusions}
We have presented a self-consistent analysis of the effects of MCP emission (via on-shell and off-shell plasmon decay) on the TRGB by implementing this model into a modified version of \mesa. As discussed in Section~\ref{sec:eloss}, we have implemented the total energy loss through both the transverse and longitudinal plasmon decay channels, using the in-medium matrix element for the corresponding decay process, as well as the full spectral function for the longitudinal and transverse plasmons. The decay of the longitudinal mode, which we included for the first time, is the dominant contribution to energy loss for higher MCP masses and lower stellar temperatures. Using these rates, we simulate stars starting with a pre-main-sequence model and evolve until the TRGB. We find that excess energy loss from MCP emission can alter the density and temperature profiles of the stellar interior, lowering the energy loss rates from SM processes (e.g. plasmon decay to neutrinos), which partly compensates for the effect of MCP emission. We also find that the emission morphology of MCPs is quite distinct from SM emission channels, occurring outside of the degenerate helium core. The net effect is that if MCPs exist, the TRGB should look substantially brighter due to their emission. We have checked that the size of this effect is robust to varying assumptions about stellar ages and compositions, mass-loss, and convective mixing. To set a limit on the existence of MCPs from the TRGB, we use 15 GCs whose parallax distances are measured with \emph{Gaia}. Our resulting limit is stronger and more robust against theoretical uncertainties compared to previous analyses that made simplifying assumptions that are not supported by our simulations. 

The analysis presented in this work paves the way for several avenues that may further improve the limits on MCPs. When setting limits on MCPs from the horizontal branch, previous analyses did not self-consistently account for deviations in the stellar density and temperature profiles compared to the SM-only stellar structure. Given that we see such deviations in our simulations, it will be worthwhile to evolve forward in time to see the progression and to determine whether the true MCP emission can be enhanced or quenched compared to the MCP emission assuming a SM-only stellar structure. Futhermore, we have not accounted for any trapping effects due to magnetic fields; such effects were previously estimated to be relevant in the Sun for large $q$, potentially leading to non-trivial energy transport~\cite{Vinyoles:2015aba}. It will therefore be worthwhile to consider magnetic trapping for post-main-sequence stars, though little remains known when it comes to magnetic fields in stellar interiors. Finally, there may be an opportunity to set stellar bounds on MCPs from distinct stellar observables with different systematic uncertainties, including asteroseismology. Other stellar populations could provide additional information. For instance, younger clusters may provide more stringent information, as the ``plateau'' in the TRGB luminosity versus stellar mass does not dip as quickly as in the SM when MCPs are included. We leave consideration of all these directions to future work.

\acknowledgements
We thank 
Aldo Serenelli for helpful discussions. AF, QL and ACV are supported by the Arthur B. McDonald Canadian Astroparticle Physics Research Institute, NSERC and the Ontario Government through an Early Researcher Award. SH was supported in part by a Trottier Space Institute Fellowship. SH and KS acknowledge support from a Natural Sciences and Engineering Research Council of Canada Subatomic Physics Discovery Grant and from the Canada Research Chairs program. VM was supported by a Mitacs Globalink Research Internship. Computing equipment was funded by the Canada Foundation for Innovation and the Ontario Government. Research at Perimeter Institute is supported by the Government of Canada through the Department of Innovation, Science, and Economic Development, and by the Province of Ontario.

\bibliography{LEDBH.bib}
\onecolumngrid
\newpage 
\appendix
\section{Comparison of bounds computed with parallax and ZAHB distances}
\label{app}

\begin{figure}
\includegraphics[width=0.45\textwidth]{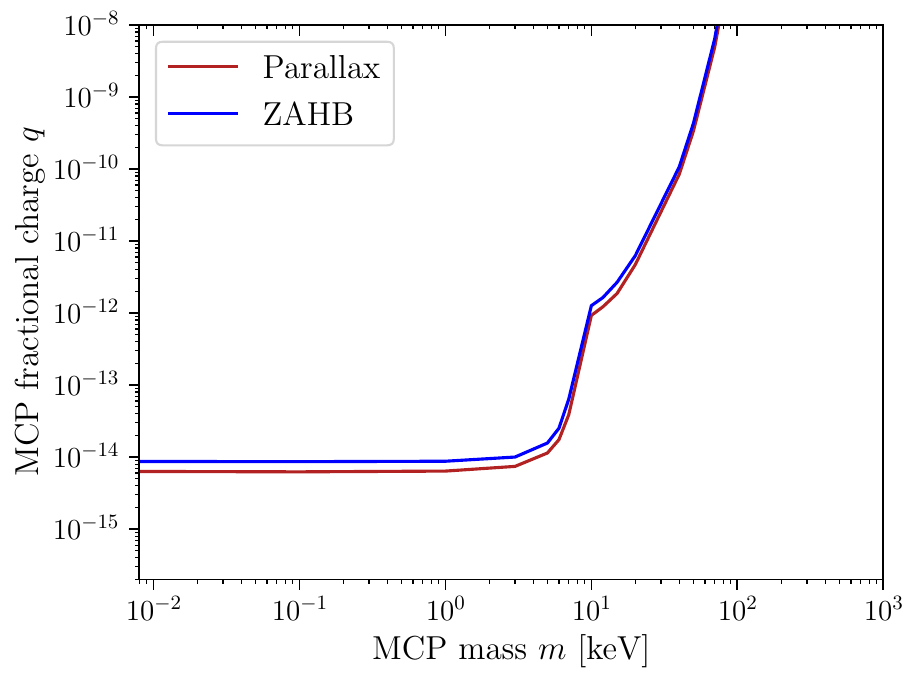}
    \caption{Same as Fig.~\ref{fig:constraint} with constraints constructed from the ZAHB results of 18 GCs.}
    \label{fig:constraint_ZAHB}
\end{figure}

We have computed equivalent bounds as the ones presented in Fig.~\ref{fig:constraint} but using bolometric magnitudes inferred from ZAHB distance calibrations instead. These comprise 18 of the 22 GCs presented in Ref. \cite{Straniero:2020iyi}, chosen to span the same range of metallicities covered by our simulations used for the main results. These  are shown in Fig. \ref{fig:constraint_ZAHB}. They are slightly weaker than constraints obtained using the parallax distance determinations, plateauing at $q < 10^{-14}$ at low MCP masses. Table \ref{tab:constraints} provides numerical values for each value of $m$ simulated in this work. 
We keep the parallax results as our recommended set of constraints, for the reasons stated in the main body: these parallax distance measurements are less susceptible to new physics effects, and the distance uncertainties can be more readily interpreted in a statistical sense.

\begin{table}[t]
    \centering
    \begin{tabular}{c|c c}
     \hline
     \hline  
{$m\,$[keV]}& {$\log_{10} q$ [parallax] } & {$\log_{10} q$ [ZAHB]} 
 \\ [0.5ex] \hline
0.01 & -14.20 & -14.06 \\
0.1 & -14.20 & -14.06 \\
1.0 & -14.19 & -14.06 \\
3.0 & -14.13 & -14.00 \\
5.0 & -13.94 & -13.80 \\
6.0 & -13.76 & -13.60 \\
7.0 & -13.41 & -13.19 \\
10.0 & -12.03 & -11.90 \\
12.0 & -11.91 & -11.78 \\
15.0 & -11.73 & -11.57 \\
20.0 & -11.33 & -11.20 \\
40.0 & -10.07 & -9.97 \\
50.0 & -9.47 & -9.37 \\
70.0 & -8.30 & -8.19 \\
100.0 & -6.60 & -6.43 \\
 \hline \hline      
    \end{tabular}
    \caption{95\% CL constraints on the MCP charge $q$ from the data set of \cite{Straniero:2020iyi}, using parallax distance determinations, shown in Fig.~\ref{fig:constraint}, and ZAHB distance determinations. These are compared in Fig.~\ref{fig:constraint_ZAHB}.} 
    \label{tab:constraints}
\end{table}

\end{document}